\documentclass[nohyper,12pt,letterpaper]{JHEP3}
\usepackage{amsfonts,amssymb,amsmath,graphicx}
\newcommand{\be}{\begin{equation}}
\newcommand{\ee}{\end{equation}}
\newcommand{\ben}{\begin{displaymath}}
\newcommand{\een}{\end{displaymath}}
\newcommand{\bea}{\begin{eqnarray}}
\newcommand{\eea}{\end{eqnarray}}
\newcommand{\bean}{\begin{eqnarray*}}
\newcommand{\eean}{\end{eqnarray*}}

\def\l {\lambda}

\def\g {\gamma}

\def\s {\sigma}

\newcommand{\commentout}[1]{}

\newcommand{\beq}{\begin{equation}}
\newcommand{\eeq}{\end{equation}}
\newcommand{\beqr}{\begin{displaymath}}
\newcommand{\eeqr}{\end{displaymath}}
\newcommand{\beqa}{\begin{eqnarray}}
\newcommand{\eeqa}{\end{eqnarray}}
\newcommand{\beqar}{\begin{eqnarray*}}
\newcommand{\eeqar}{\end{eqnarray*}}

\title{\LARGE Generating AdS String Solutions}

\author{Antal Jevicki, Kewang Jin, Chrysostomos Kalousios, Anastasia Volovich\\
Department of Physics, Brown University, Box 1843, Providence, RI 02912 \\
E-mail: \email{antal, jin, ckalousi, nastja@het.brown.edu}}

\abstract{
We use a Pohlmeyer type reduction to generate classical string
solutions in $AdS$ spacetime. In this framework we describe a correspondence
between spikes in $AdS_3$ and soliton profiles of the sinh-Gordon equation.
The null cusp string solution and its closed spinning string
counterpart are related to the sinh-Gordon vacuum. We construct classical
string solutions corresponding to sinh-Gordon solitons, antisolitons and
breathers by the inverse scattering technique. The breather solutions
can also be reproduced by the sigma model dressing method.}

\keywords{Classical string solutions, AdS/CFT, integrable systems}

\preprint{\tt{BROWN-HET-1492}}

\begin{document}

\section{Introduction}

Classical string solutions in $AdS_5 \times S^5$ have provided a lot of data in exploring various aspects of the AdS/CFT correspondence (see \cite{GKP,Tseytlin:2004cj,Tseytlin:2004xa,Plefka:2005bk} for review). Recently Alday and Maldacena gave a prescription for computing gluon scattering amplitudes using AdS/CFT \cite{Alday:2007hr}.
The prescription is equivalent to finding a classical string solution with boundary conditions determined by the gluon momenta. The value of the scattering amplitude is then related to the area of this solution. Using this prescription and the solution originally constructed in \cite{Kruczenski:2002fb} they found agreement with the conjectured iteration relations for perturbative multiloop amplitudes for four gluons \cite{Anastasiou:2003kj, Bern:2005iz, Bern:2006ew, Cachazo:2006az, Cachazo:2007ad}. Several recent papers \cite{Jevicki:2007pk, Mironov:2007qq, Astefanesei:2007bk, Kruczenski:2007cy} have studied various aspects of the classical string solutions (see \cite{Ito:2007zy}--\cite{Itoyama:2007ue} for other developments). For the case of four and five gluons the results are fixed by dual conformal symmetry \cite{Alday:2007he, Drummond:2007au}. For a large number of gluons the amplitude at strong coupling was computed in \cite{Alday:2007he} and it disagreed with the corresponding limit of the gauge theory guess \cite{Bern:2005iz}. In order to test the multiloop iterative structure of gauge theory amplitudes it would be very important to construct the string solution for six gluons and more.

Classical string theory on $R \times S^2$ (or $R \times S^3$) is equivalent to classical sine-Gordon theory (or complex sine-Gordon theory) via Pohlmeyer reduction \cite{Pohlmeyer:1975nb}. De Vega and Sanchez showed that similarly string theory on $AdS_2$, $AdS_3$ and $AdS_4$ is equivalent to Liouville theory, sinh-Gordon theory and $B_2$ Toda theories respectively \cite{DeVega:1992xc, deVega:1992yz, Barbashov:1980kz, Larsen:1996gn, Combes:1993rw}. Moreover, very recently a sine-Gordon-like action has been proposed for the full Green-Schwarz superstring in $AdS_5 \times S^5$ \cite{Grigoriev:2007bu, mikhailov}. Classical solitons in both theories should be in one to one correspondence. Indeed, giant magnon solutions on $R \times S^2$ and $R \times S^3$ map to one soliton solution in sine-Gordon and complex sine-Gordon respectively \cite{Hofman:2006xt,Chen:2006gea}.

Integrability of string theory on $AdS_5 \times S^5$ allows the use of algebraic methods to construct solutions of the nonlinear equations of motion. Given a vacuum solution of an integrable nonlinear equation, the dressing method provides a way to construct a new solution which also satisfies the equations of motion by using an associated linear system \cite{Zakharov:1973pp,Harnad:1983we}. In \cite{Spradlin:2006wk, Kalousios:2006xy} the dressing method was used to construct classical string solutions describing scattering and bound states of magnons on $R \times S^5$ and various subsectors, such as $R \times S^2$ and $R \times S^3$, by dressing the vacuum corresponding to a pointlike string moving around the equator of the sphere at the speed of light. In \cite{Ishizeki:2007kh} it was used to construct solutions describing the scattering of spiky strings on a sphere \cite{spikyS5} by starting with a different vacuum, a static string wrapped around the equator of the sphere.

In \cite{Jevicki:2007pk} the applicability of the dressing method to the problem of finding Euclidean minimal area worldsheets in AdS was demonstrated. We took as a vacuum the null cusp string solution constructed in \cite{Kruczenski:2002fb} (which was later generalized and given a new interpretation in \cite{Alday:2007hr}). We dressed this vacuum and found new minimal area surfaces in $AdS_3$ and $AdS_5$. These solutions generically trace out timelike curves on the boundary, and might be relevant to studies of the propagation of massive particles in gauge theory. The vacuum solution \cite{Alday:2007hr, Kruczenski:2002fb} can be related by analytic continuation and a conformal transformation to a closed string energy eigenstate (an infinite string limit of GKP string \cite{GKP, Kruczenski:2007cy}). In this paper we outline the dressing method for Minkowskian worldsheets in AdS and construct new string solutions by starting with an infinite
closed spinning string. We also show that the spikes of the long GKP string can be mapped
to sinh-Gordon solitons at the boundary of AdS.

We use the inverse scattering method to construct string solutions corresponding to sinh-Gordon solitons, antisolitons, breathers and soliton scattering solutions. The sigma model solutions can be constructed in terms of wavefunctions of the Pohlmeyer reduced model \footnote{A different solution generating technique based on Pohlmeyer-type reduction was employed for string solutions on $AdS_3 \times S^1$ in \cite{Hayashi:2007bq}.} \cite{Neveu:1977cr}. The advantage of this method is that it allows us to construct a string solution starting from any sinh-Gordon solution. All one has to do is to solve a linear system with coefficients depending on the chosen sinh-Gordon solution. Notice that in the dressing method one is also solving a linear system, but the difference is that in the dressing method the coefficients of the system depend on the chosen vacuum solution of the string equations, whereas in this method the coefficients depend only on the sinh-Gordon or reduced system solution.
This is advantageous because any sinh-Gordon solution is generaly simpler than the corresponding
sigma model solution.

The paper is organized as follows. In section two we review the Pohlmeyer reduction and inverse scattering method for constructing string solutions from sinh-Gordon solutions. In section three various sinh-Gordon solutions are reviewed. In section four explicit string solutions  are constructed and the physical meanings are discussed. It would be interesting to understand the physics of these new string solutions better. In section five we reproduce the breather solutions by the dressing method.

\section{Pohlmeyer reduction for AdS strings}

In this section we review the Pohlmeyer reduction for string theory in $AdS_d$ space following \cite{DeVega:1992xc}. We also review how to write down string solutions in terms of the wavefunctions of the
sinh-Gordon inverse problem \cite{Neveu:1977cr}.

We parameterize $AdS_d$ with $d+1$ embedding coordinates $\vec{Y}$ subject to the constraint
\begin{equation}
\vec{Y} \cdot \vec{Y} \equiv -Y_{-1}^2-Y_0^2+Y_1^2+Y_2^2+ \dots +Y_{d-1}^2=-1.
\end{equation}
The conformal gauge equation of motion for strings in $AdS_d$ is
\begin{equation}
\partial \bar{\partial} \vec{Y}-(\partial \vec{Y} \cdot \bar{\partial} \vec{Y}) \vec{Y}=0
\label{eom}
\end{equation}
subject to the Virasoro constraints
\begin{equation}
\partial \vec{Y} \cdot \partial \vec{Y}=\bar{\partial} \vec{Y} \cdot \bar{\partial} \vec{Y}=0.
\label{vir}
\end{equation}
Here we use coordinates $z$ and $\bar{z}$ related to Minkowski worldsheet coordinates $\tau$ and $\sigma$ by $z={1\over2}(\sigma-\tau),~\bar{z}={1\over2}(\sigma+\tau)$ with $\partial=\partial_{\sigma}-\partial_{\tau},~\bar{\partial}=\partial_{\sigma}+\partial_{\tau}$.

Now let us show the equivalence of the string equations (\ref{eom},~\ref{vir}) to the generalized sinh-Gordon model. To make the reduction we first choose a basis
\be
e_i=(\vec{Y},~\bar{\partial} \vec{Y},~\partial \vec{Y}, ~\vec{B}_4, \cdots, ~\vec{B}_{d+1}),
\label{basis}
\ee
where $i=1,2 \cdots d+1$ and the vectors $\vec{B}_{k}$ with $k=4, 5 \cdots d+1$ are orthonormal
\be
\vec{B}_k \cdot \vec{B}_l=\delta_{kl},~~~\vec{B}_k \cdot \vec{Y}=\vec{B}_k \cdot \partial \vec{Y}=\vec{B}_k \cdot \bar{\partial} \vec{Y}=0.
\ee

Defining
\begin{eqnarray} \label{map}
\alpha &\equiv& \alpha(z,\bar{z})=\ln (\partial \vec{Y} \cdot \bar{\partial}\vec{Y}),\\
u_k &\equiv& u_k(z,\bar{z})=\vec{B}_k \cdot \bar{\partial}^2 \vec{Y},\\
v_k &\equiv& v_k(z,\bar{z})=\vec{B}_k \cdot \partial^2 \vec{Y},
\end{eqnarray}
where $k=4,5 \cdots d+1$, the equation of motion for $\alpha$ becomes
\begin{equation}
\partial \bar{\partial} \alpha-e^\alpha-e^{-\alpha} \sum_{i=4}^{d+1} u_i v_i=0.
\label{sinhgordon}
\end{equation}
This is called the generalized sinh-Gordon model. We can find the evolution of the vectors $u_i$ and $v_i$ by expressing the derivatives of the basis (\ref{basis}) in terms of the basis itself. In $d=2$, $u=v=0$ and the equation (\ref{sinhgordon}) becomes the Liouville equation. In $d=3$ and $d=4$ it can be reduced to sinh-Gordon and $B_2$ Toda models respectively \cite{DeVega:1992xc}.

Now let us discuss the $d=3$ case in more detail. For the case of $AdS_3$, one can write an explicit formula for $\vec{B}_4$
\be
B_{4a} \equiv e^{-\alpha} \epsilon_{abcd} ~Y_b ~\partial Y_c ~\bar{\partial} Y_d,
\ee
where $a,b,c,d=1,2,3,4$ and $\epsilon_{abcd}$ is the antisymmetric Levi-Civita tensor. The equations of motion can then be rewritten as
\be
\bar{\partial} e_i=A_{ij}(z,\bar{z}) e_j,\hspace{.2in} \partial e_i=B_{ij}(z,\bar{z}) e_j,
\label{ABmat}
\ee
where
\be
A=\begin{pmatrix} 0 & 1 & 0 & 0 \cr 0 & \bar{\partial}\alpha & 0 & u \cr e^\alpha & 0 & 0 & 0 \cr 0 & 0 & -ue^{-\alpha} & 0 \end{pmatrix},\hspace{.2in} B=\begin{pmatrix} 0 & 0 & 1 & 0 \cr e^\alpha & 0 & 0 & 0 \cr 0 & 0 & \partial \alpha & v \cr 0 & -ve^{-\alpha} & 0 & 0 \end{pmatrix}.
\ee
The integrability condition $\partial A-\bar{\partial} B+[A,B]=0$ implies $u=u(\bar{z})$, $v=v(z)$ and $\partial \bar{\partial} \alpha-e^\alpha-uv e^{-\alpha}=0$. We can make a change of variables
\be
\alpha(z,\bar{z})=\hat{\alpha}(z,\bar{z})+\frac{1}{2} \ln(-u(\bar{z}) v(z))
\ee
to bring the equation (\ref{sinhgordon}) to a standard sinh-Gordon form
\be
\partial \bar{\partial} \hat{\alpha} -4\sinh\hat{\alpha}=0.
\label{shg}
\ee

\subsection{Constructing string solutions from sinh-Gordon solutions}
In this section we use the Pohlmeyer reduction to express solutions of the equations (\ref{eom}, \ref{vir}) in terms of solutions of the sinh-Gordon equation (\ref{sinhgordon}) \cite{Neveu:1977cr}. The idea is to first rewrite the matrices $A_{ij}$ and $B_{ij}$ which appear in (\ref{ABmat}) in a manifestly $SO(2,2)$ symmetric way. Then recalling that $SO(2,2)$ is isomorphic to $SU(1,1) \times SU(1,1)$ one can expand $A_{ij}$ and $B_{ij}$ in terms of $SU(1,1)$ generators. Defining
\begin{equation}
A_1=\begin{pmatrix} {-i\over 2 \sqrt{2}}(ue^{-\alpha/2}+e^{\alpha/2}) & {i\over 4}\bar{\partial}\alpha-{1\over 2 \sqrt{2}}(ue^{-\alpha/2}-e^{\alpha/2}) \cr -{i\over 4}\bar{\partial}\alpha-{1\over 2 \sqrt{2}}(ue^{-\alpha/2}-e^{\alpha/2}) & {i\over 2 \sqrt{2}}(ue^{-\alpha/2}+e^{\alpha/2}) \end{pmatrix},
\end{equation}
\begin{equation}
A_2=\begin{pmatrix} {-i\over 2 \sqrt{2}}(ve^{-\alpha/2}-e^{\alpha/2}) & -{i\over 4}\partial \alpha+{1\over 2 \sqrt{2}}(ve^{-\alpha/2}+e^{\alpha/2}) \cr {i\over 4}\partial \alpha+{1\over 2 \sqrt{2}}(ve^{-\alpha/2}+e^{\alpha/2}) & {i\over 2 \sqrt{2}}(ve^{-\alpha/2}-e^{\alpha/2}) \end{pmatrix},
\end{equation}
\begin{equation}
B_1=\begin{pmatrix} {-i\over 2 \sqrt{2}}(ue^{-\alpha/2}-e^{\alpha/2}) & {i\over 4}\bar{\partial}\alpha-{1\over 2 \sqrt{2}}(ue^{-\alpha/2}+e^{\alpha/2}) \cr -{i\over 4}\bar{\partial}\alpha-{1\over 2 \sqrt{2}}(ue^{-\alpha/2}+e^{\alpha/2}) & {i\over 2 \sqrt{2}}(ue^{-\alpha/2}-e^{\alpha/2}) \end{pmatrix},
\end{equation}
\begin{equation}
B_2=\begin{pmatrix} {-i\over 2 \sqrt{2}}(ve^{-\alpha/2}+e^{\alpha/2}) & -{i\over 4}\partial \alpha+{1\over 2 \sqrt{2}}(ve^{-\alpha/2}-e^{\alpha/2}) \cr {i\over 4}\partial \alpha+{1\over 2 \sqrt{2}}(ve^{-\alpha/2}-e^{\alpha/2}) & {i\over 2 \sqrt{2}}(ve^{-\alpha/2}+e^{\alpha/2}) \end{pmatrix},
\end{equation}
we can rewrite equations (\ref{ABmat}) in terms of two unknown complex vectors $\phi=(\phi_1, \phi_2)^T$ and $\psi=(\psi_1, \psi_2)^T$ as
\begin{equation}
\bar{\partial} \phi=A_1\phi,\hspace{.2in} \partial \phi=A_2\phi,
\label{linsysone}
\end{equation}
\begin{equation}
\bar{\partial} \psi=B_1 \psi,\hspace{.2in} \partial \psi=B_2\psi.
\label{linsystwo}
\end{equation}
The vectors $\phi$ and $\psi$ are normalized $\phi^\dagger \phi=\phi_1^*\phi_1-\phi_2^*\phi_2=\psi^\dagger \psi=\psi_1^*\psi_1-\psi_2^*\psi_2 =1$. In other words, given a solution $\alpha(z,\bar{z}), u(\bar{z})$ and $v(z)$ of the sinh-Gordon equation, we can find $\phi$ and $\psi$ such that they solve the above linear system. Then the string solution is given by
\beqa
Z_1&\equiv&Y_{-1}+i Y_0=\phi_1^*\psi_1-\phi_2^*\psi_2, \label{stsol1} \\
Z_2&\equiv&Y_1+i Y_2=\phi_2^*\psi_1^*-\phi_1^*\psi_2^*. \label{stsol2}
\eeqa
This formula follows from the isomorphism between $SO(2,2)$ and the product of two copies of $SU(1,1)$ parametrized by the matrices $\begin{pmatrix} \phi_1 & \phi_2^* \cr \phi_2 &\phi_1^* \end{pmatrix}$ and $\begin{pmatrix}  \psi_1 &  \psi_2^* \cr \psi_2 &\psi_1^* \end{pmatrix}$.

\section{Review of sinh-Gordon solutions}
The sinh-Gordon equation (\ref{shg}) has a vacuum solution
\be
\hat{\alpha}_0=0~~~{\rm or}~~~\alpha_0=\ln 2.
\ee
The one-soliton solutions are
\be
{\alpha}_{s,\bar{s}}=\ln 2 \pm \ln \bigl(\tanh^2 \gamma (\sigma-v\tau)\bigr),
\label{solsol}
\ee
where $v$ is the velocity of the solitons and $\g=1/\sqrt{1-v^2}$.

We can also consider solutions periodic in $\sigma$
\be
{\alpha'}_{s,\bar{s}}=\ln 2 \pm \ln \bigl(\tan^2 \gamma (\sigma-v\tau)\bigr).
\ee

Multi-soliton solutions can be constructed via the B\"{a}cklund transformation. If we call the plus solution of (\ref{solsol}) soliton and the minus solution antisoliton, the two-(anti)soliton solution is given by
\be
\alpha_{ss,\bar{s}\bar{s}}=\ln 2 \pm \ln \Bigl[ {v\cosh X-\cosh T \over v\cosh X+\cosh T} \Bigr]^2,
\ee
where $X=2\g\s$, $T=2v\g \tau$, and the soliton-antisoliton solution is given by
\be
\alpha_{s\bar{s}}=\ln 2 \pm \ln \Bigl[ {v\sinh X-\sinh T \over v\sinh X+\sinh T} \Bigr]^2.
\ee
Here the solutions are in the center of mass frame with $v_1=-v_2=v$.

If we analytically continue the soliton-antisoliton solution and take $v$ to be imaginary, $v=iw$, we get the breather solution of the sinh-Gordon system
\be
\alpha_B=\ln2 \pm \ln \Bigl[{w\sinh X_B -\sin T_B \over w\sinh X_B +\sin T_B} \Bigr]^2,
\label{breather}
\ee
where $X_B=2\s/\sqrt{1+w^2}$ and $T_B=2w\tau/\sqrt{1+w^2}$. In order to make the center mass move with velocity $v_c$, one can make a boost by replacing $\sigma \to \g_c(\sigma-v_c \tau)$ and $\tau \to \g_c(\tau-v_c \sigma)$, where $\g_c=1/\sqrt{1-v_c^2}$.

\section{String solutions}

\subsection{Vacuum}
\begin{figure}
\centering
\includegraphics[width=0.9\textwidth]{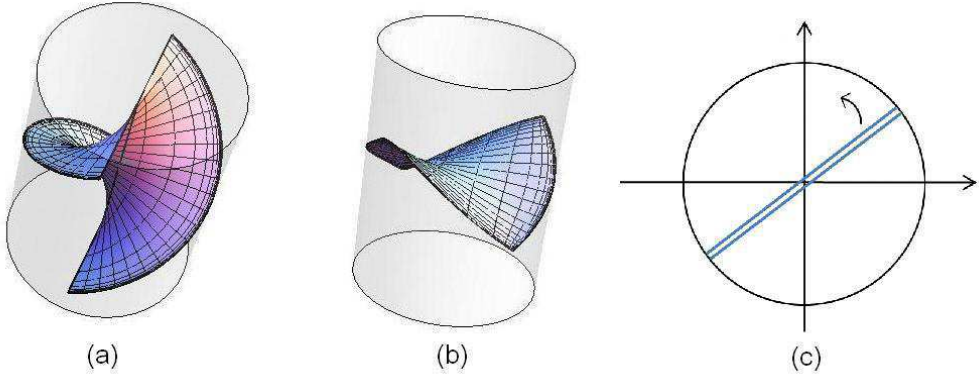}
\caption{The vacuum solution in (a) Minkowskian and (b) Euclidean worldsheet plotted in $AdS_3$ coordinates. (c) Top view of Minkowskian vacuum solution. The boundary of the worldsheet touches the boundary of $AdS$ space.}
\label{vacuum}
\end{figure}

Now let us look at some examples. Starting with the sinh-Gordon vacuum $u=2,v=-2,\alpha_0=\ln2$, the results of solving the linear system (\ref{linsysone},~\ref{linsystwo}) are
\be
\phi_1=e^{-i\tau} \hspace{.2in} \phi_2=0 \hspace{.2in} \psi_1=\cosh\sigma \hspace{.2in} \psi_2=-\sinh\sigma.
\ee
Then the Minkowskian worldsheet solution is given by (see fig. \ref{vacuum})
\beqa
Z_1&=&e^{i\tau}\cosh\sigma,\label{stvacuum1} \\
Z_2&=&e^{i\tau}\sinh\sigma.\label{stvacuum2}
\eeqa
This is the infinite string limit of spinning string \cite{GKP}.

The Euclidean worldsheet solution is obtained by making the change $\tau \to -i \tau$. Then $Y_0$ and $Y_2$ become imaginary, thus effectively exchanging places. The Euclidean vacuum solution reads
\begin{equation}
\vec{Y}_E=\begin{pmatrix} \cosh \sigma \cosh \tau \cr \sinh \sigma \sinh \tau \cr \sinh \sigma \cosh \tau \cr \cosh \sigma \sinh \tau \end{pmatrix}.
\end{equation}
This is the solution found in \cite{Kruczenski:2002fb} which was used by the authors of \cite{Alday:2007hr} to calculate the scattering amplitude for four gluons.

The energy and angular momentum can be calculated after we introduce the cutoff $\Lambda \gg 0$,
\bea
E&=&{\sqrt{\lambda} \over \pi}\int_{-\Lambda}^\Lambda d\sigma \cosh^2 \sigma \approx {\sqrt{\lambda} \over 4 \pi}e^{2\Lambda},\\
S&=&{\sqrt{\lambda} \over \pi}\int_{-\Lambda}^\Lambda d\sigma \sinh^2 \sigma \approx {\sqrt{\lambda} \over 4 \pi}e^{2\Lambda},\\
E-S&=&{\sqrt{\lambda} \over \pi}\int_{-\Lambda}^{\Lambda} d\sigma \sim {\sqrt{\lambda} \over \pi} \ln {4 \pi \over \sqrt{\lambda}}S,
\eea
which is exactly the result of \cite{GKP}.

\subsection{Long strings in $AdS_3$ as sinh-Gordon solitons}
Consider the GKP spinning string solution found in \cite{GKP}
\beqa
Z_1&=&e^{i\tau}\cosh \rho(\sigma),\\
Z_2&=&e^{i\omega\tau}\sinh\rho(\sigma),
\eeqa
where
\be
\rho(\sigma)=am(i \sigma|1-\omega^2)
\label{GKPrho},
\ee
and $am$ the Jacobi amplitude function. In the infinite string limit $\omega \to 1$
this solution reduces to (\ref{stvacuum1}, \ref{stvacuum2}).  The corresponding sinh-Gordon solution is given by
\begin{equation}
\alpha=\ln(2{\rho^\prime}^2)=\ln\left(2dn^2 (i \sigma|1-\omega^2)\right),
\end{equation}
where $dn$ is the Jacobi elliptic function.

Taking the (\ref{GKPrho}) solution ${\rho^\prime}^2=\cosh^2\rho-\omega^2\sinh^2\rho$ we can expand $\rho$ near one of spikes (turning points of the string) and let $\omega=1+2\eta$, where $\eta \ll 1$, to get
\be
{\rho^\prime}^2 \sim e^{2\rho}(e^{-2\rho}-\eta).
\ee
Denoting $u=e^{-\rho}$ the above equation becomes
\be
{u^\prime}^2 \sim u^2-\eta.
\ee
If we choose the location of the spike to be at $\sigma=\sigma_0$, we find
\be
\rho(\sigma)=-\ln \bigl(\sqrt{\eta} \cosh (\sigma-\sigma_0) \bigr).
\ee
Now we can use the map (\ref{map}) to find the sinh-Gordon solution corresponding to this spinning string
\be
\alpha=\ln(2{\rho^\prime}^2)=\ln(2\tanh^2(\sigma-\sigma_0))
\ee
This is exactly the one-soliton solution to the sinh-Gordon equation (\ref{sinhgordon}). Therefore, the long string limit of the spinning string solution \cite{GKP} itself is a two-soliton configuration of the sinh-Gordon system and the solitons are located near the boundary of AdS.

\subsection{One-soliton solutions}
\begin{figure}[h!]
\begin{center}
\includegraphics[width=0.7\textwidth]{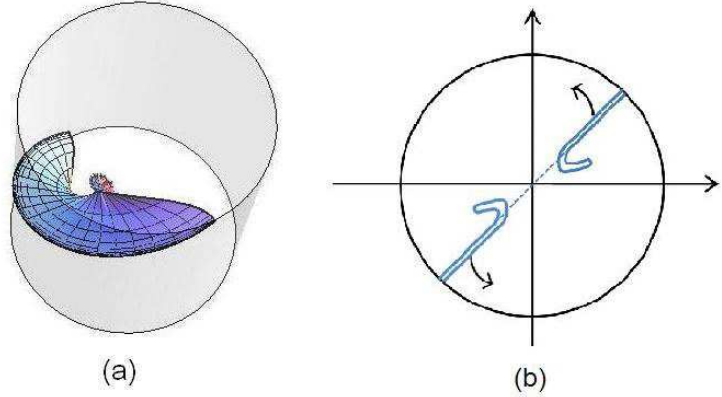}
\caption{The one-soliton solution in (a) Minkowskian worldsheet plotted in $AdS_3$ coordinates. (b) Top view of the Minkowskian one-soliton solution. Please note the curvature of the string changes with the evolution of time.}
\label{soliton}
\end{center}
\end{figure}

Let us describe the method of constructing string solutions corresponding to one-soliton sinh-Gordon solution in detail. Start with the sinh-Gordon solution
\begin{equation}
\alpha_s=\ln2+\ln(\tanh^2 \sigma).
\end{equation}
The matrices entering into the linear system (\ref{linsysone},~\ref{linsystwo}) are given by
\bea
A_1 &=& \begin{pmatrix} -i\coth 2\sigma & (i-1) \text{csch} \hspace{.02in} 2\sigma \cr -(i+1) \text{csch} \hspace{.02in} 2\sigma & i \coth 2\sigma \end{pmatrix}, \\
A_2 &=& \begin{pmatrix} i\coth 2\sigma & -(i+1) \text{csch} \hspace{.02in} 2\sigma \cr (i-1) \text{csch} \hspace{.02in} 2\sigma & -i \coth 2\sigma \end{pmatrix}, \\
B_1 &=& \begin{pmatrix} -i \text{csch} \hspace{.02in} 2\sigma & i \text{csch} \hspace{.02in} 2\sigma-\coth 2\sigma \cr -i \text{csch} \hspace{.02in} 2\sigma - \coth 2\sigma & i \text{csch} 2\sigma \end{pmatrix}, \\
B_2 &=& \begin{pmatrix} i \text{csch} \hspace{.02in} 2\sigma & -i \text{csch} \hspace{.02in} 2\sigma-\coth 2\sigma \cr i \text{csch} \hspace{.02in} 2\sigma - \coth 2\sigma & -i \text{csch} \hspace{.02in} 2\sigma \end{pmatrix}.
\eea
The spinors that solve the linear system are
\bea
\phi_1 &=& e^{-i\tau}\cosh ({1\over2} \ln \tanh \sigma), \\
\phi_2 &=& -e^{-i\tau}\sinh ({1\over2} \ln \tanh \sigma), \\
\psi_1 &=& (\tau+i) \cosh({1\over2} \ln \sinh 2\sigma)-\tau \sinh ({1\over2} \ln \sinh 2\sigma), \\
\psi_2 &=& -(\tau+i) \sinh({1\over2} \ln \sinh 2\sigma)+\tau \cosh ({1\over2} \ln \sinh 2\sigma).
\eea
Then we use (\ref{stsol1},~\ref{stsol2}) to find the corresponding string solution (see fig. \ref{soliton})
\begin{eqnarray}
Z_{1}^s &=& {e^{i\tau} \over 2\sqrt{2} \cosh \sigma}\bigl(2\tau+i(\cosh 2\sigma+2) \bigr),\\
Z_{2}^s &=& {e^{i\tau} \over 2\sqrt{2} \cosh \sigma}\bigl(-2\tau-i\cosh 2\sigma \bigr).
\end{eqnarray}
Because of the Lorentz invariance, we can always boost the solution as $\sigma \rightarrow \g(\sigma-v \tau ), \tau \rightarrow \g(\tau-v\sigma) $. Notice this differs from the magnon case, where the boost symmetry of sine-Gordon translates into a non-obvious symmetry on the string side \cite{Hofman:2006xt}.

The Euclidean worldsheet solution is obtained by making the changes $\tau \to -i \tau$. Then $Y_{-1}$ and $Y_1$ become imaginary, thus effectively exchanging places. The Euclidean one-soliton solution reads
\be
\vec{Y}_E^s={1\over 2\sqrt{2} \cosh \sigma}\begin{pmatrix} {2 \tau \cosh \tau-\sinh \tau \cosh 2 \sigma} \cr { -2 \tau \sinh \tau + \cosh \tau (\cosh 2\sigma+2)} \cr { -2\tau\cosh \tau+\sinh \tau(\cosh 2\sigma+2)} \cr {2\tau \sinh \tau-\cosh \tau \cosh 2\sigma} \end{pmatrix}.
\ee

One can easily compute the energy and angular momentum
\be
E=\int_{-\Lambda}^\Lambda d\sigma {\sqrt{\lambda} \over 16\pi \cosh^2 \sigma}(1+8\tau^2+4\cosh 2\sigma +\cosh 4\sigma) \approx {\sqrt{\lambda} \over \pi}({1\over8} e^{2\Lambda}+\tau^2),
\ee
\be
S=\int_{-\Lambda}^\Lambda d\sigma {\sqrt{\lambda} \over 16\pi \cosh^2 \sigma}(1+8\tau^2-4\cosh 2\sigma +\cosh 4\sigma) \approx {\sqrt{\lambda} \over \pi}({1\over8} e^{2\Lambda}+\tau^2).
\ee
If we neglect the $\tau$ dependence since the exponential term is much larger than the square term, we have
\be
E-S=\int_{-\Lambda}^{\Lambda} {\sqrt{\lambda} \over 2\pi} \cosh 2\sigma \hspace{.02in} \text{sech}^2 \hspace{.02in} \sigma d\sigma \sim {\sqrt{\lambda} \over \pi}\ln {8\pi \over \sqrt{\lambda}}S.
\ee
The energy is not conserved because there is momentum flow at the asymptotic end of the string and the string itself is not closed.

Similarly, the one-antisoliton string solution corresponding to $\alpha_{\bar{s}}$ is given by
\beqa
Z_{1}^{\bar{s}}&=&{e^{i\tau} \over  2\sqrt{2} \sinh \sigma}\bigl( 2\tau-i\cosh 2\sigma \bigr),\\
Z_{2}^{\bar{s}}&=&{e^{i\tau} \over 2\sqrt{2} \sinh \sigma}\bigl(-2\tau+i(\cosh 2\sigma-2)\bigr),
\eeqa
whereas the periodic in $\sigma$ string solutions mapping to $\alpha_s^\prime$ and $\alpha_{\bar{s}}^\prime$ are respectively
\be
\vec{Y}'_s={1\over 2\sqrt{2} \cos \sigma}\begin{pmatrix} { 2\tau\cosh\tau-\sinh\tau\cos 2\sigma  } \cr { 2\tau\sinh\tau-\cosh\tau(\cos 2\sigma+2)} \cr { 2\tau\cosh\tau-\sinh\tau(\cos 2\sigma+2) } \cr {-2\tau\sinh\tau+\cosh\tau\cos 2\sigma} \end{pmatrix},
\ee
\be
\vec{Y}'_{\bar{s}}={1 \over 2\sqrt{2} \sin \sigma} \begin{pmatrix} { 2\tau\cosh\tau+\sinh\tau\cos 2\sigma} \cr { 2\tau\sinh\tau+\cosh\tau(\cos 2\sigma-2)} \cr { -2\tau\cosh\tau-\sinh\tau(\cos 2\sigma-2) } \cr {2\tau\sinh\tau+\cosh\tau\cos 2\sigma} \end{pmatrix}.
\ee
Energy and angular momentum are singular for those solutions.

\subsection{Two-soliton solutions}
For the two-soliton solution $\alpha_{ss}$ in sinh-Gordon, the spinors are
\begin{eqnarray}
\phi_1 &=& e^{i\tau}{i\sqrt{1-v^2}\sinh T+iv\sinh T \over \sqrt{\cosh^2 T-v^2\cosh^2 X}}, \\
\phi_2 &=& e^{i\tau}{v\sinh X \over \sqrt{\cosh^2 T-v^2\cosh^2 X}},\\
\psi_1 &=& {(\sqrt{1-v^2}\cosh X+i\sinh T)\cosh \sigma -\sinh X\sinh \sigma \over \sqrt{\cosh^2 T-v^2\cosh^2 X}},\\
\psi_2 &=& {(-\sqrt{1-v^2}\cosh X+i\sinh T)\sinh \sigma +\sinh X\cosh \sigma \over \sqrt{\cosh^2 T-v^2\cosh^2 X}},
\end{eqnarray}
where $X=2\g\s$, $T=2v\g \tau$. The two-soliton string solution is \footnote{We occasionally use the notation $\text{sh}$ and $\text{ch}$ for $\sinh$ and $\cosh$ to simplify otherwise lengthy formulas.}
\begin{eqnarray}
Z_1^{ss} &=& e^{-i\tau}{v\text{ch} T \text{ch} \sigma +\text{ch} X\text{ch} \sigma-\sqrt{1-v^2}\text{sh} X \text{sh} \sigma+i\sqrt{1-v^2}\text{sh} T \text{ch}\sigma \over \text{ch} T +v\text{ch} X}, \\
Z_2^{ss} &=& e^{-i\tau}{v\text{ch} T \text{sh} \sigma +\text{ch} X\text{sh} \sigma-\sqrt{1-v^2}\text{sh} X \text{ch} \sigma+i\sqrt{1-v^2}\text{sh} T \text{sh}\sigma \over \text{ch} T +v\text{ch} X}.
\end{eqnarray}

\begin{figure}[h!]
\begin{center}
\includegraphics[width=0.7\textwidth]{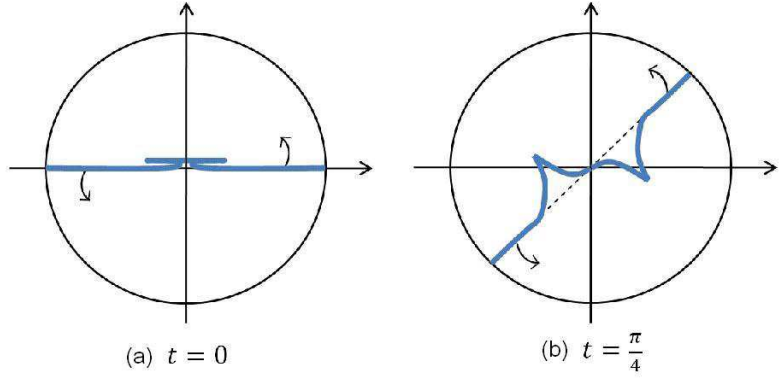}
\caption{The Minkowskian two-soliton solution with $v={1 \over \sqrt{5}}$ at different global time (a) $t=0$, (b) $t=\pi /4$.}
\label{twosoliton}
\end{center}
\end{figure}

Fig. \ref{twosoliton} shows the shape of the two-soliton string at two different global time instants. In fig. \ref{twosoliton}(a), the string is folded along the $x$ axis, whereas in fig. \ref{twosoliton}(b), we find the usual bulk spikes.

The two-soliton solution can also be anallytically continued to the Euclidean worldsheet under the $\tau \to -i\tau$ change.  Then $Y_0$ and $Y_2$ become imaginary and they effectively change place.

The two-antisoliton string solution can be constructed in the same way and it only differs
from the two-soliton solution by three signs, the second and third terms in the numerator and the second term in the denominator.

For the soliton-antisoliton $\alpha_{s\bar{s}}$ solution, the result is
\begin{eqnarray}
Z_1^{s\bar{s}} &=& e^{-i\tau}{v\text{sh} T \text{ch} \sigma \pm \text{sh} X\text{ch} \sigma \mp \sqrt{1-v^2}\text{ch} X \text{sh} \sigma+i\sqrt{1-v^2}\text{ch} T \text{ch}\sigma \over \text{sh} T \pm v\text{sh} X}, \\
Z_2^{s\bar{s}} &=& e^{-i\tau}{v\text{sh} T \text{sh} \sigma \pm \text{sh} X\text{sh} \sigma \mp \sqrt{1-v^2}\text{ch} X \text{ch} \sigma+i\sqrt{1-v^2}\text{ch} T \text{sh}\sigma\over \text{sh} T \pm v\text{sh} X}.
\end{eqnarray}

Finally, we take the breather solution of sinh-Gordon (\ref{breather}) and we solve the spinors from (\ref{linsysone},~\ref{linsystwo}) to find the string solution
\begin{multline}
Z_1^B = {e^{-i\tau} \over \sin T_B \pm w\text{sh} X_B} \bigl\{-w\sin T_B \text{sh} \sigma \pm \text{sh} X_B\text{sh} \sigma \\
\mp \sqrt{1+w^2}\text{ch} X_B \text{ch} \sigma) +i\sqrt{1+w^2}\cos T_B \text{sh}\sigma \bigr\},
\label{stbreather1}
\end{multline}
\begin{multline}
Z_2^B = {e^{-i\tau} \over \sin T_B \pm w\text{sh} X_B} \bigl\{-w\sin T_B \text{ch} \sigma \pm \text{sh} X_B\text{ch} \sigma \\
\mp \sqrt{1+w^2}\text{ch} X_B \text{sh} \sigma) +i\sqrt{1+w^2}\cos T_B \text{ch}\sigma \bigr\},
\label{stbreather2}
\end{multline}
where $X_B=2\s/\sqrt{1+w^2}$ and $T_B=2w\tau/\sqrt{1+w^2}$.

\section{$AdS$ dressing method}
The dressing method allows the construction of solutions to nonlinear classically integrable equations. Many of the equations here are similar to \cite{Jevicki:2007pk} and the reader may look there for further details. Here we use the dressing method to construct new string theory solutions on $AdS_3$ for a Minkowskian worldsheet.

We recast the system (\ref{eom},~\ref{vir}) into the form of a principal $SU(1,1)$ chiral model for the matrix-valued field $g(z,\bar{z})$ that satisfies the equation of motion
\begin{equation}
\bar{\partial}A+\partial\bar{A}=0,
\label{eomdr}
\end{equation}
where the currents $A$ and $\bar{A}$ are given by
\begin{eqnarray}
A&=&i\partial g g^{-1}\,,\\
\bar{A}&=&i\bar{\partial}g g^{-1}.
\end{eqnarray}
As an example we can consider the $AdS_3$ case and easily prove the equivalence of equations (\ref{eomdr}) to equations (\ref{eom}, \ref{vir}) using the following $SU(1,1)$ parametrization
\begin{equation}
g=\left( \begin{array}{cc} Y_{-1}+i Y_0 & Y_1+i Y_2\\Y_1-i Y_2 & Y_{-1}-i Y_0\\ \end{array}\right)
\label{param}
\end{equation}
that satisfies
\begin{equation}
g^{\dag}Mg=M, \qquad M=\left( \begin{array}{cc} 1&0\\0&-1 \end{array} \right), \qquad \text{det }g=1.
\end{equation}

The second order system (\ref{eomdr}) is equivalent to the first order system
\begin{equation}
i\partial \Psi=\frac{A \Psi}{1-\lambda}, \qquad i\bar{\partial}\Psi=\frac{\bar{A}\Psi}{1+\lambda}
\label{drlin}
\end{equation}
for the auxiliary field $\Psi(z,\bar{z},\lambda)$.  The complex number $\lambda$ is called the spectral parameter.

In order to apply the dressing method we start with any known solution that we call the vacuum and we solve (\ref{drlin}) to find $\Psi(\lambda)$ subject to the condition
\begin{equation}\label{psicondition}
\Psi(\lambda=0)=g.
\end{equation}
Since we want $\Psi(\lambda)$ to be an $SU(1,1)$ element we further impose the unitarity constraint
\begin{equation}
\Psi^\dag(\bar{\lambda})M\Psi(\lambda)=M
\end{equation}
and demand that
\begin{equation}
\text{det }\Psi(\lambda)=1.
\end{equation}
Furthermore we consider the transformation
\begin{equation}
\Psi^{\prime}(\lambda)=\chi(\lambda)\Psi(\lambda)
\end{equation}
and seek a $\chi(\lambda)$, the dressing factor, that depends on $z$ and $\bar{z}$ in such a way that $\Psi^{\prime}(\lambda)$ still satisfies (\ref{drlin}). In that case $\Psi^{\prime}(\lambda=0)$ is a new solution to (\ref{eomdr}).

For the $AdS_3$ case we can take the dressing factor to be
\begin{equation}
\chi(\lambda)=I+\frac{\lambda_1-\bar{\lambda}_1}{\lambda-\lambda_1} P,
\end{equation}
where $\lambda_1$ is an arbitrary complex number and the projector $P$ is given by
\begin{equation}
P=\frac{\upsilon_1\upsilon_1^{\dagger}M}{\upsilon_1^{\dagger} M\upsilon_1}, \quad \upsilon_1=\Psi(\bar{\lambda})e,
\end{equation}
where $e$ is an arbitrary vector with constant complex entries called the polarization vector.  The projector $P$ does not depend on the length of the $e$ vector.

The determinant of $\chi(\lambda)$ is $\bar{\lambda}_1/\lambda_1$ and if we want our solution to sit in $SU(1,1)$ we should rescale $\chi(\lambda)$ by the compensating factor $\sqrt{\frac{\lambda_1}{\bar{\lambda}_1}}$.

Putting everything together the new solution $g^{\prime}=\Psi^{\prime}(\lambda=0)$ to the system (\ref{eom}, \ref{vir}) is given by
\begin{equation}
g^{\prime}=\sqrt{\frac{\lambda_1}{\bar{\lambda}_1}}\left(I+\frac{\lambda_1-\bar{\lambda}_1}{-\lambda_1} P \right)g.
\end{equation}

\subsection{Breather solution}
Here we apply the above dressing method to dress the vacuum in order to find new string theory solutions in $AdS_3$. As a vacuum we choose the solution (\ref{stvacuum1}, \ref{stvacuum2}). Using the $AdS_3$ parametrization (\ref{param}) we find that the currents $A$, $\bar{A}$ are given by
\begin{eqnarray}
A&=&\left(\begin{array}{cc} 1 & i e^{2 i \tau}\\ i e^{-2 i \tau} & -1 \end{array}\right), \\
\bar{A}&=&\left(\begin{array}{cc} -1 & i e^{2 i \tau}\\ i e^{-2 i \tau} & 1 \end{array}\right).
\end{eqnarray}
Then a solution to the system (\ref{drlin}) subject to the unitarity constraints yields
\begin{equation}
\Psi(\lambda)=
\left(
  \begin{array}{cc}
    e^{i \tau}\left(\cosh Z-\frac{i\lambda \sinh Z}{\sqrt{1-\lambda^2}}\right) & \frac{e^{i \tau}\sinh Z}{\sqrt{1-\lambda^2}} \\
    \frac{e^{-i \tau}\sinh Z}{\sqrt{1-\lambda^2}} & e^{-i \tau}\left(\cosh Z+\frac{i\lambda \sinh Z}{\sqrt{1-\lambda^2}}\right) \\
  \end{array}
\right),
\end{equation}
where
\begin{equation}
Z=z\left(\frac{1+\lambda}{1-\lambda}\right)^{1/2}+\bar{z}\left(\frac{1-\lambda}{1+\lambda}\right)^{1/2}.
\end{equation}

The general solution, that can be read off from the components of the matrix field $g^{\prime}=\chi g$ in terms of the polarization vector $e$ is rather complicated, so we present here the full solution in the case of $e=(1~~i)$.  The dressed solution is
\begin{eqnarray}\label{solution}
Y_{-1}+i Y_0&=& e^{i \tau}\frac{N_1}{D},\\
Y_{1}+i Y_2&=& e^{i \tau}\frac{N_2}{D},
\end{eqnarray}
where
\begin{equation}
\begin{split}
N_1&=\sqrt{1-\lambda_1^2}\cosh Z_1 ((\bar{\l}_1-\l_1)\sqrt{1-\bar{\l}_1^2} \cosh \bar{Z}_1 (\sinh \s-i \cosh \s)\\
&\quad \quad -(\bar{\l}_1-1)\sinh \bar{Z}_1 ((\l_1+\bar{\l}_1) \cosh \s-i (\l_1-\bar{\l}_1) \sinh \s))\\
&\quad +(\lambda_1-1)\sinh Z_1 ((\l_1-\bar{\l}_1)(\bar{\l}_1-1)\sinh \bar{Z}_1(i\cosh \s+\sinh \s)\\
&\quad \quad +\sqrt{1-\bar{\l}_1^2}\cosh \bar{Z}_1((\l_1+\bar{\l}_1) \cosh \s+i (\l_1-\bar{\l}_1) \sinh \s)),
\end{split}
\end{equation}
\begin{equation}
\begin{split}
N_2&=\sqrt{1-\lambda_1^2}\cosh Z_1 ((\bar{\l}_1-\l_1)\sqrt{1-\bar{\l}_1^2} \cosh \bar{Z}_1 (\cosh \s-i \sinh \s)\\
&\quad \quad +i(\bar{\l}_1-1)\sinh \bar{Z}_1 ((\l_1-\bar{\l}_1) \cosh \s+i (\l_1+\bar{\l}_1) \sinh \s))\\
&\quad +(\lambda_1-1)\sinh Z_1 ((\l_1-\bar{\l}_1)(\bar{\l}_1-1)\sinh \bar{Z}_1(\cosh \s+i\sinh \s)\\
&\quad \quad +\sqrt{1-\bar{\l}_1^2}\cosh \bar{Z}_1(i(\l_1-\bar{\l}_1) \cosh \s+ (\l_1+\bar{\l}_1) \sinh \s)),
\end{split}
\end{equation}
\begin{equation}
    D=2|\lambda_1| \left((\lambda_1-1)\sqrt{1-\bar{\lambda}_1^2}\cosh \bar{Z}_1 \sinh Z_1-\sqrt{1-\lambda_1^2}(\bar{\lambda}_1-1)\cosh Z_1\sinh \bar{Z}_1 \right),
\end{equation}
where \footnote{$Z_1$, $\bar{Z}_1$ should not to be confused with the embedding string coordinates in (\ref{stsol1}, \ref{stsol2}).}
\begin{eqnarray}
Z_1&=&z\left(\frac{1+\lambda_1}{1-\lambda_1}\right)^{1/2}+\bar{z}\left(\frac{1-\lambda_1}{1+\lambda_1}\right)^{1/2},\\
\bar{Z}_1&=&z\left(\frac{1+\bar{\lambda}_1}{1-\bar{\lambda}_1}\right)^{1/2}+\bar{z}\left(\frac{1-\bar{\lambda}_1}{1+\bar{\lambda}_1}\right)^{1/2},
\end{eqnarray}

This is precisely the same solution (\ref{stbreather1}, \ref{stbreather2}) that we obtained in the previous section using the inverse scattering method as we can easily see by expressing the spectral parameter $\lambda_1$ in terms of center mass velocity $v_1$ and the frequency $w_1$ of the breather solution by
\begin{equation}
\lambda_1=\frac{w_1-iv_1}{w_1 v_1-i}.
\end{equation}

\section*{Acknowledgments}

We are grateful to M. Abbott, I. Aniceto, M. Spradlin for comments and discussions. This work is supported by DOE grant DE-FG02-91ER40688. The research of AV is also supported by NSF CAREER Award PHY-0643150.

\appendix
\section{Conventions}

Here we summarize the standard conventions for global $AdS_3$ that we have used in preparing the figures. We parameterize the SU(1,1) group element  as
\begin{eqnarray*}
Z_1&=&e^{i t} \sec \theta,\\
Z_2&=&e^{i \phi} \tan \theta,
\end{eqnarray*}
where $t$ is the global time, $\phi$ the azimuthal angle, and $\theta$ runs from $0$ in the interior of the $AdS_3$ cylinder to $\pi/2$ at the boundary of $AdS_3$. In terms of these quantities the parametric plots in the figures have Cartesian coordinates
\begin{equation*}
(x,y,z) = ( \theta \cos \phi, \theta \sin \phi, t)
\end{equation*}
and the boundary of $AdS_3$ is the cylinder $x^2 + y^2 = (\pi/2)^2$.

\end{document}